\documentclass[aps,pra,groupedaddress,twocolumn,amsmath]{revtex4-1}

\usepackage{graphicx}
\usepackage{subfigure}
\usepackage{color}
\usepackage{cases}
\usepackage{float}
\usepackage{hyperref}
\usepackage{epstopdf}
\usepackage{natbib}
\begin{document}

\title[Analytical and numerical studies of Bose-Fermi mixtures in a one-dimensional harmonic trap]{Analytical and numerical studies of Bose-Fermi mixtures in a one-dimensional harmonic trap}

\author{A.~S. Dehkharghani, F.~F. Bellotti, and N.~T. Zinner}
\affiliation{Department of Physics and Astronomy, Aarhus University, DK-8000 Aarhus C, Denmark}

\begin{abstract}
In this paper we study a mixed system of bosons and fermions
with up to six particles in total. All particles 
are assumed to have the same mass. The two-body interactions 
are repulsive and are assumed to have equal strength in 
both the Bose-Bose and the Fermi-Boson channels. The particles
are confined externally by a harmonic oscillator one-body
potential. For the case of four particles, two identical 
fermions and two identical bosons, we focus on the 
strongly interacting regime and analyze the system using
both an analytical approach and DMRG calculations using a 
discrete version of the underlying continuum Hamiltonian. 
This provides us with insight into both the ground state
and the manifold of excited states that are almost degenerate
for large interaction strength. Our results show great variation 
in the density profiles for bosons and fermions in different 
states for strongly 
interacting mixtures.
By moving to slightly larger systems, we find that the ground state of balanced
mixtures of four to six particles tends to separate bosons and fermions
for strong (repulsive) interactions.
On the other hand, in imbalanced Bose-Fermi mixtures we find pronounced odd-even effects in systems of five particles. 
These few-body results suggest that question of phase separation in one-dimensional 
confined mixtures are very sensitive to system composition, both for 
the ground state and the excited states.
\end{abstract}




\maketitle

\section{Introduction}

One-dimensional quantum systems have been some of the most widely used examples in the theoretical physics literature \cite{lieb2013mathematical,abramowitz1965,baxter2016exactly,giamarchi2003quantum,plischke2006equilibrium}, mostly due to their simplicity in contrast to higher dimensions. Moreover, recent experimental realizations with cold atoms in one dimension (1D) \cite{blochRMP2008,lewensteinAiP2007,esslingerARoCMP2010,moritzPRL2003,
stoferlePRL2003, KinoshitaS2004, KinoshitaN2006, ParedesN2004, HallerS2009, haller2010pinning, PaganoNP2014,onofrio2016} have made it possible to test new and old theories for such systems. In recent years, experimental manipulations have evolved so rapidly and in many new ways that one can now build a Fermi sea one atom at a time in low-dimensions \cite{SerwaneS2011, zurnPRL2012,zurnPRL2013,WenzS2013,MurmannPRL2015a,
MurmannPRL2015b} and therefore be able to study and understand the transition between few- and many-body systems \cite{ZinnerEWoC2016, HofmannPRA2016}. Other existing predictions such as some from the Bethe ansatz \cite{Bethe1931, YangPRL1967, OelkersJPA2006, HaoPRA2006} and that of the Tonks-Girardeau gas \cite{TonksPR1936, GirardeauJoMP1960} have also been studied and tested successfully \cite{KinoshitaS2004, ParedesN2004, HallerS2009, GuanRMP2013, VingSR2016}. Even mixing different kinds of particles and species with tunable interaction between the atoms is now feasible \cite{MurmannPRL2015a,Spethmann2012} and therefore opening up for developing and testing new kinds of theories.

Despite being simpler than higher dimensions, 1D systems are sometimes more interesting and exotic \cite{zurnPRL2012}. This is mainly due to the 1D nature that prohibits any exchange of particles without their wave function overlapping with each other. At the same time, this allows one to build a chain of atoms with one or several internal components and start manipulating them by changing the surroundings \cite{zurnPRL2012,ZuernPRL2013} or tuning the interactions between the particles through Feshbach Resonances \cite{chinRMP2010, OlshaniiPRL1998}. In the long term, this can be used for transport of quantum information \cite{Endres1024} and 
such knowledge may be advantageous in a variety of technological applications such as nano-tubes and nano-wires\cite{Altomare2013}.

At the theoretical level 1D systems have been intensively studied both analytically and numerically in the recent years \cite{SowinskiPRA2013, DeuretzbacherPRA2014,VolosnievPRA2015, HuNJoP2016, YangPRA2016, PecakNJoP2016, YangPRA2015, CampbellPRA2014, Garcia-MarchNJoP2014, Garcia-MarchPRA2015, Garcis-MarchJoPBAMaOP2016, ZoellnerPRA2008, TempfliNJoP2009, BrouzosPRL2012, BrouzosPRA2013, DeuretzbacherPRL2008, SowinskiEEL2015, zinnerJoPGNaPP2013, LoftJoPBAMaOP2016, LoftCPC2016, DehkharghaniSR2015, GirardeauPRA2004, GirardeauPRL2007, LevinsenSA2015,DeuretzbacherPRA2016, KollerPRL2016, ZinnerPRA2015}. Numerically, the DMRG method has been one of the most successful methods. It was firstly developed for discrete Hamiltonians \cite{WhitePRL1992, SchollwockAP2011, FuehringerAdP2008, WallNJoP2012} and then pushed to the limit of continuous systems even for strongly interacting particles \cite{FangPRA2011,HuNJoP2016a}. Analytically, new and different kinds of methods have been developed and are in use as well \cite{AndersenAe2015, VolosnievNC2014, VolosnievFS2014, DehkharghaniJPB2016}. However, different kind of methods for all regimes have both advantages and disadvantages, and they must be used with care \cite{BellottiPJD2017}.

Here we study both numerical DMRG methods and analytical exact solutions of Bose-Fermi mixtures in a one-dimensional harmonic trap in the limit of strong interaction, and we compare how well the DMRG method captures the physics of the system. In addition, we compare our results with other recent studies of such systems \cite{FangPRA2011, HuNJoP2016a,deuretzbacher2016spin, DasPRL2003, CazalillaPRL2003, ImambekovPRA2006, WangPRA2012}. In particular, the study of Deuretzbacher {\it et al.}~\cite{deuretzbacher2016spin} have recently considered a spin-chain model approach to Bose-Fermi mixtures, which is similar to the analytical approach to the strongly interacting limit that we use below. Our analytical results are in agreement with the work in \cite{deuretzbacher2016spin}.

The structure of the paper is as follows. In Section \ref{sec.sm} we present our systems and the methods. Boson-fermion mixtures are discussed in Section \ref{sec.mss} in connection with fermionization of such systems in the strongly interacting limit. Concluding remarks are given in Section \ref{sec.conc}.

\section{System and methods} \label{sec.sm}
Our system consists of a mixture of bosons and fermions with
total number of particles $N=N_b+N_f$, where $N_b$ denotes the number of identical bosons and $N_f$ is the number of identical fermions. All the particles have the same mass $m$ and are confined with the same trap frequency $\omega$ in a one-dimensional harmonic trap, $V(x)=m\omega^2x^2/2$. 
The two-body interaction is short-range and we model it by the zero-range model as a Dirac $\delta$-function. Furthermore, 
we assume that the interactions are purely repulsive throughout this paper.  
The full Hamiltonian can therefore be written as
\begin{equation} 
\mathcal{H}_c=\sum_{i=1}^{N} \left( -\frac{\hbar^2 }{2 m} \frac{\partial^2}{\partial x_i^2} + V(x_i) \right) + \sum_{i<j}^{N} U_{ij}(x_i-x_j) ,
\label{hamil}
\end{equation}
with the interaction matrix $U_{ij}(x_i-x_j)=g_{ij} \delta(x_i-x_j)$ and $g_{ij}>0$. 
The interaction strength is assumed to be the same $g_{ij}=g$ if particle $i$ and particle $j$ are either from different species or they are identical bosons with self-interaction. 
However, due to the Pauli exclusion principle we may set $g_{ij}=0$ when $i$ and $j$ are fermions as 
identical fermions will not feel any effect of a zero-range potential.

In order to study the system using DMRG in the usual discrete form, we need to specify a lattice model for the 
Hamiltonian above. Here we take discrete lattice Hamiltonian of the form
\begin{widetext}
\begin{eqnarray} 
\mathcal{H}_d & =  -t \sum_{j=1}^{N-1} \left( b_{j}^{\dagger} b_{j+1} + b_{j+1}^{\dagger} b_{j} \right) 
-t \sum_{j=1}^{N-1} \left( f_{j}^{\dagger} f_{j+1} + f_{j+1}^{\dagger} f_{j} \right) 
+ U_{bf} \sum_{j=1}^{N} n_{b,j} n_{f,j} & \nonumber\\
 &+ \frac{U_{bb}}{2} \sum_{j=1}^{N} n_{b,j} \left(n_{b,j}-1 \right) 
+ V_h \sum_{j=1}^{N} (j-L/2)^2 \left(n_{b,j}+n_{f,j} \right) & , 
\label{hub}
\end{eqnarray} 
\end{widetext}
where $b_j$ and $f_j$ are the bosonic and fermionic field operators, respectively, acting on a site $j$ and with the corresponding density operators $n_{b,j}=b_{j}^{\dagger} b_{j}$ and $n_{f,j}=f_{j}^{\dagger} f_{j}$. The tunneling constant, $t$, is the equivalent of the kinetic term in the continuous case, while $U_{bf}$ and $U_{bb}$ are the on-site interactions. The strengths of the on-site interactions are correspondingly assumed to be the same, $U_{bf}=U_{bb}=U$, and the strength of the harmonic potential is 
called $V_h$. In the low-density limit where the number of particles is much less than the number of discrete lattice sites, this model should reproduce the physics of the continuous system.

In the limit of very strong interaction strengths an exact analytical wave function has been derived \cite{VolosnievNC2014,VolosnievFS2014,DecampAe2016} for the continuous Hamiltonian in Eq.~\eqref{hamil}. This method makes use of the fact that up to linear order in $1/g$ one can obtain the exact slope of the eigenstates and use this fact to derive an analytical solution for each eigenstate. The method is applicable for any arbitrary confining geometries and in a particular case with a harmonic trap. The case of 
four particles with two-component fermions has been treated and derived in details recently \cite{BellottiPJD2017}. The main result here is that the wave function of the two-component mixture system with $N=N_b+N_f$ particles can be written as a non-trivial combination of the antisymmetric product of the first $N$ eigenstates to the non-interacting part of Eq.~\eqref{hamil}
\begin{widetext}
\begin{numcases}
{\psi(x_1,...,x_N)= \label{psiana}}
a_1 \Psi_A & for $x_{b_1}<...<x_{b_{N_b}}<x_{f_1}<...<x_{f_{N_f}}$ $(b...bf...f)$ \nonumber \\
a_2 \Psi_A & for $x_{b_1}<...<x_{f_1}<x_{b_{N_b}}<...<x_{f_{N_f}}$ $(b...fb...f)$ \nonumber \\
\vdots & $\vdots$ \\
a_M \Psi_A & for $x_{f_{N_f}}<...<x_{f_1}<x_{b_{N_b}}<...<x_{b_1}$ $(f...fb...b)$ \nonumber 
\end{numcases}
\end{widetext}
where $M=N!/(N_b! N_f!)$ denotes the number of degeneracy at $1/g\to0$ and $x_n$ is the coordinate of the $n^{th}$ particle. The coefficients, $\{a_1, a_2, \dots, a_M \}$ are then found by considering the slope of the energy as one approaches $1/g\to0$. \\

To solve the lattice model we will use DMRG as stated above.
In recent years, many libraries that implement the DMRG method \cite{WallNJoP2012, Wall2009,itensor} have been developed to solve discretized lattice Hamiltonians. In this article we make use of two independently developed open-source codes, one from L. D. Carr and his group \cite{WallNJoP2012, Wall2009} and the other from the iTensor project \cite{itensor}. We find 
consistency of the results from both codes. 

A mapping between the discrete and continuous Hamiltonian has also been investigated and described recently \cite{BellottiPJD2017}. One of the important mappings here is the connection between the interaction strength $g$ and the on-site interaction $U$, which is found to be the following
\begin{equation}
U=0.10291\cdot g
\end{equation}
The above relation is found by investigating the relation between the energy calculated from the discrete Hamiltonian 
Eq.~\eqref{hub}, labeled $E_d$, and the one from the continuous Hamiltonian Eq.~\eqref{hamil}, labeled $E_c$, which is given by
\begin{equation}
\frac{E_c}{\hbar \omega}-N \frac{1}{2}=\frac{E_{d}}{\hbar \omega_{d}} -N\frac{ E_{d_{\mathrm{1p}}}}{\hbar \omega_{d}}.
\label{eshift}
\end{equation} 
Here $E_{d_\mathrm{1p}}$ is the energy one obtains when using the discrete model to solve the one particle system in a harmonic trap and $\hbar \omega_d$ is the energy difference between the ground and 1st excited states. $\hbar \omega_d$ is important to calculate in order to know the units of the discrete model. $E_c$ is the energy obtained by the continuous model and $N$ is the total number of particles. The second term on both sides subtracts the non-interacting ground state energy in the same units as the first term.

With this mapping one can compare the DMRG results with the exact solution in the strongly interacting regime. Since the exact solution is only valid in the $1/g\rightarrow 0$ limit, one can use other methods, such as variational \cite{AndersenAe2015} or exact diagonalization \cite{DehkharghaniSR2015,PecakNJoP2016} to compare the results for intermediate values of $g$. However, in this work we only focus on the DMRG results for the strong interacting regime in a mixture of bosons and fermions and how these results compare to the ones obtained by the exact methods. In addition, we are also interested in how the particles mix with the opposite species.

\begin{figure*}[!htb]%
\includegraphics[width=\linewidth]{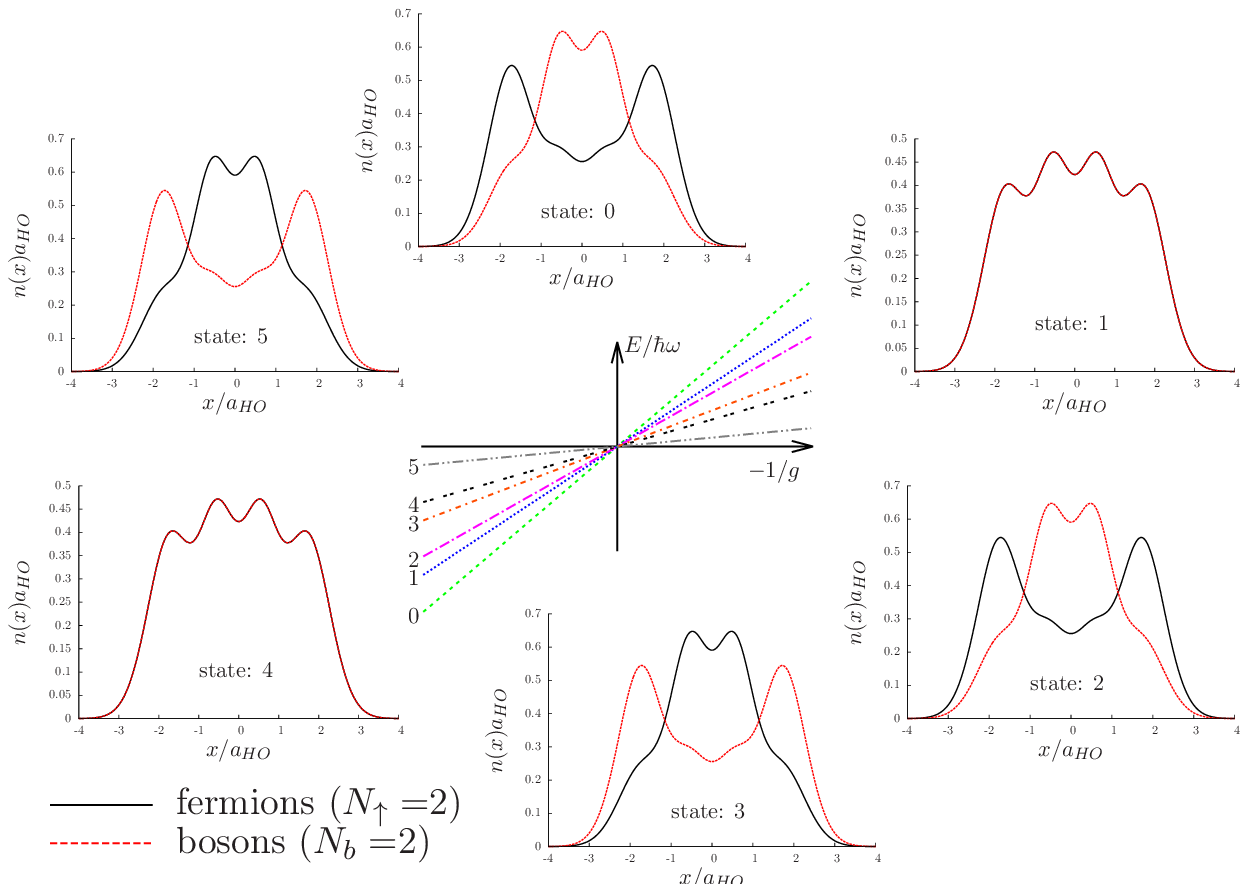}%
\caption{Energy slopes for the 6 degenerated states at $1/g=0$ and density profiles for the six states for a mixture of 2 fermions and 2 bosons. Results obtained with the wave function Eq.~\ref{psiana} following the procedure from \cite{VolosnievNC2014}.}%
\label{d22}%
\end{figure*} 

\section{Mixed species systems} \label{sec.mss}
In the following we study mixtures of identical bosons (denoted by $b$) and identical fermions (for concreteness we take them to have spin up) \cite{GirardeauPRA2004,GirardeauPRL2007}. We first consider the case study of a four-body
mixture of equal numbers of bosons and fermions. Then we proceed to study systems with 
five and six particles in order to generalize our findings and investigate what we may inferred 
about larger systems from our few-body approach.

\subsection{Two bosons and two fermions}
First we take $N_\uparrow=N_b=2$. All particles are assumed to interact with a zero-range potential of strength $g$. Due to the 
Pauli principle the fermionic particles are antisymmetric under inter-particle exchange and are therefore not affected by this zero-range interaction.
The degenerate manifold of states for strong interaction will have $M=4!/(2!2!)=6$ states. The mixture system is of particular interest since there have been speculations that the 
system will fermionize in a trivial way so that all particles will behave as essentially identical 
fermions \cite{FangPRA2011}. In Ref.~\cite{FangPRA2011} this conclusion was supported by both analytical and DMRG calculations. However, one may use the techniques of Ref.~\cite{VolosnievNC2014} and detailed calculations of Ref.~\cite{BellottiPJD2017} for the two-species system to show that the ground state does not fermionize in 
this manner. 

In figure~\ref{d22} we show the six states for the two bosons and two 
fermions mixture in the strongly interacting regime. In the 
center of the figure we display the slopes of the energy close to $1/g\to 0$, 
while along the edge of the figure we show the density of profiles of the 
six states enumerated in accordance with the slopes. The fermion density
is shown as a solid (black) line while the bosons are show by a dashed (red) 
line.
The first observation to make is that the ground state (denoted state 0) 
has a very specific manner in which it fermionizes; the two bosons are seen 
to be pushed to the center of the trap while the two fermions are left to occupy 
the edges. This may be intuitively understood from a tendency for the fermions 
to want to avoid each other, but we caution that since all the particles have 
strong two-body repulsive interactions, it is a subtle issue. The bosons could 
also be said to want to avoid each other as much as possible. In the competition 
of these effects, we see that the Pauli principle for the fermions seem to imply 
a preference for bosons in the middle and fermions out to the sides. 

A particularly peculiar second observation, is that there is an excited state (denoted
state 3), where the trend of the ground state is exactly reversed, i.e. bosons go 
the sides while fermions go to the middle. Keep in mind that in the strict limit
where $1/g=0$, all states are energetically degenerate, again pointing out the 
very subtle dynamics that determines the spatial configurations. A similar story
is seen to happen when comparing states 2 and 5 in figure~\ref{d22}, exact opposite
trends of spatial distributions of bosons and fermions are witnessed. The 
final two states denoted 1 and 4 are seen to have completely 'democratic' fermionization, 
i.e. the density profiles look exactly like a system of four identical fermions. 
Note that even though these two states look identical, their four-body wave functions
are distinct and they are orthogonal states due to different signs of different
configurations in the wave functions. In a density plot this is not visible as it 
depends on the absolute value squared of the wave functions. In comparison 
to the ground state discussed in Ref.~\cite{FangPRA2011} which has a density profile 
like states 1 and 4 in our figure~\ref{d22} and which was supported by DMRG 
evidence in \cite{FangPRA2011}, 
we speculate that this state may have been an excited state that the 
DMRG could not resolve due to the (quasi)-degenerate nature of the spectrum 
for very strong interactions which can make DMRG calculations unreliable 
\cite{BellottiPJD2017}.

The example in figure~\ref{d22} highlights how the spatial structure of 
the system is very different for different states in the strongly interacting
regime. It indicates that if one can find a way to selectively populate the 
excited states of the system, one would have access to preparation of few-body
systems with distinct quantum spatial structure, and in turn to tailoring of 
quantum magnetism of mixed systems in small systems. 

\begin{figure*}[!htb]%
\includegraphics[width=0.9\linewidth]{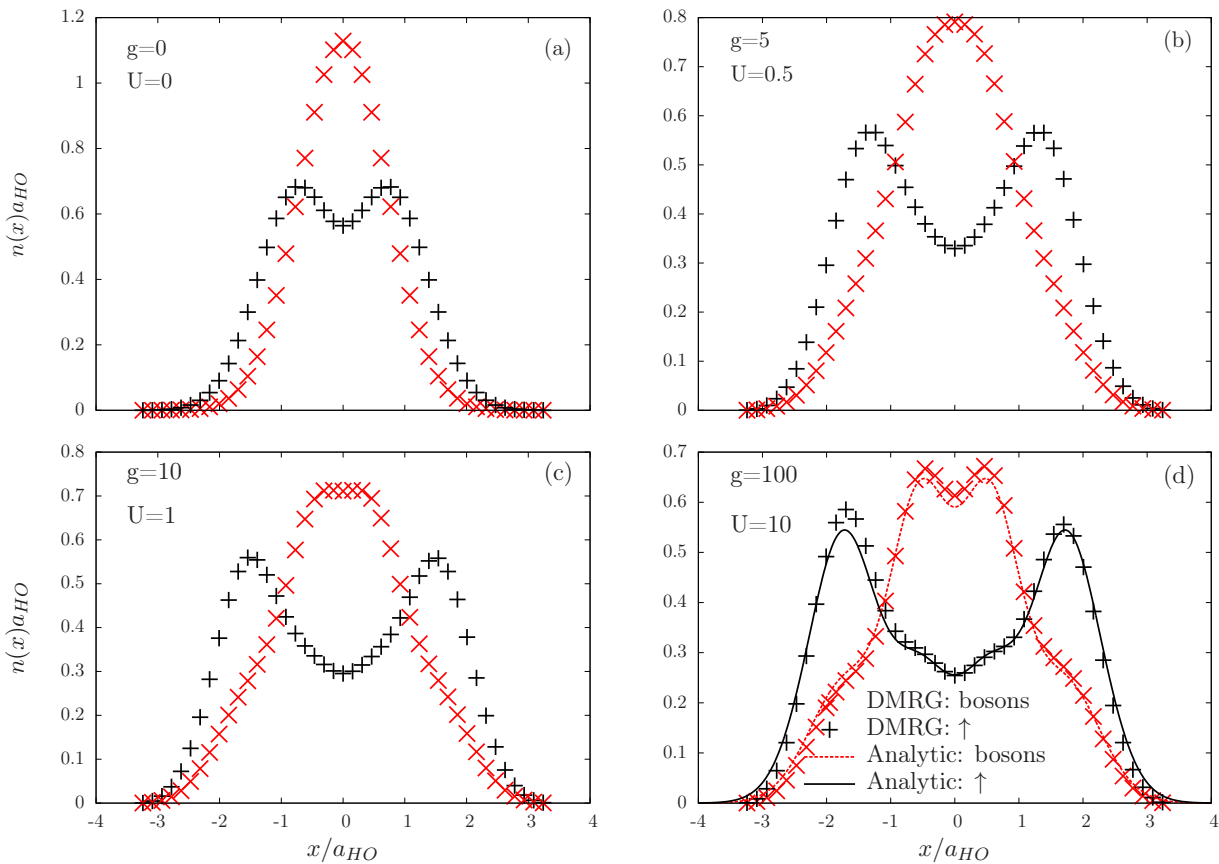}%
\caption{DMRG results for the evolution of the ground state 
density profile as function of the interaction parameter $g$ ($U=0.10291 g$). Lines on the last panel are the exact solution in the strongly interacting limit Eq.~\eqref{psiana}.
The parameters of the DMRG calculations for the Hamiltonian in Eq.~\eqref{hub} are
$V_h/t=7\cdot 10^{-6}$, $t=1$, $L=128$, and $U_{bf}=U_{bb}=U$.}%
\label{gs22}%
\end{figure*}

In order to compare the analytical results of figure~\ref{d22}, we perform 
DMRG calculations and show the density profiles in figure~\ref{gs22}. 
Indeed, one can get to the strongly interacting regime and reproduce the 
analytical profiles, but this must be done with care. Figure~\ref{gs22}
shows the evolution of the density profiles as we go from the 
non-interacting, $g=U=0$, to the strongly interaction regime 
where $g\sim 100$ and in turn $U\sim 10$. All the points in the 
figure have been calculated with DMRG while the solid lines in 
figure~\ref{gs22}(d) are from the analytical calculations of the 
density profiles as discussed above. We do see a bit of 
deviation from the analytical profiles but not alarmingly so. 
This does show clearly the need for analytical insights in this 
regime.

In order to further analyze the four-body mixture, we focus on the
spatial configurations in more detail. The six states that become 
degenerate at $1/g=0$ are superpositions of the $M=6$ independent ways
to order the two bosons and two fermions on a line. The amplitude 
of a given configuration of the four bodies is highly non-trivial 
and the weights must be determined by solving for the coefficients
$a_k$ in Eq.~\eqref{psiana}. This can be done in the manner outlined 
in \cite{VolosnievNC2014}. 

\begin{figure*}[!htb]%
\includegraphics[width=0.7\linewidth]{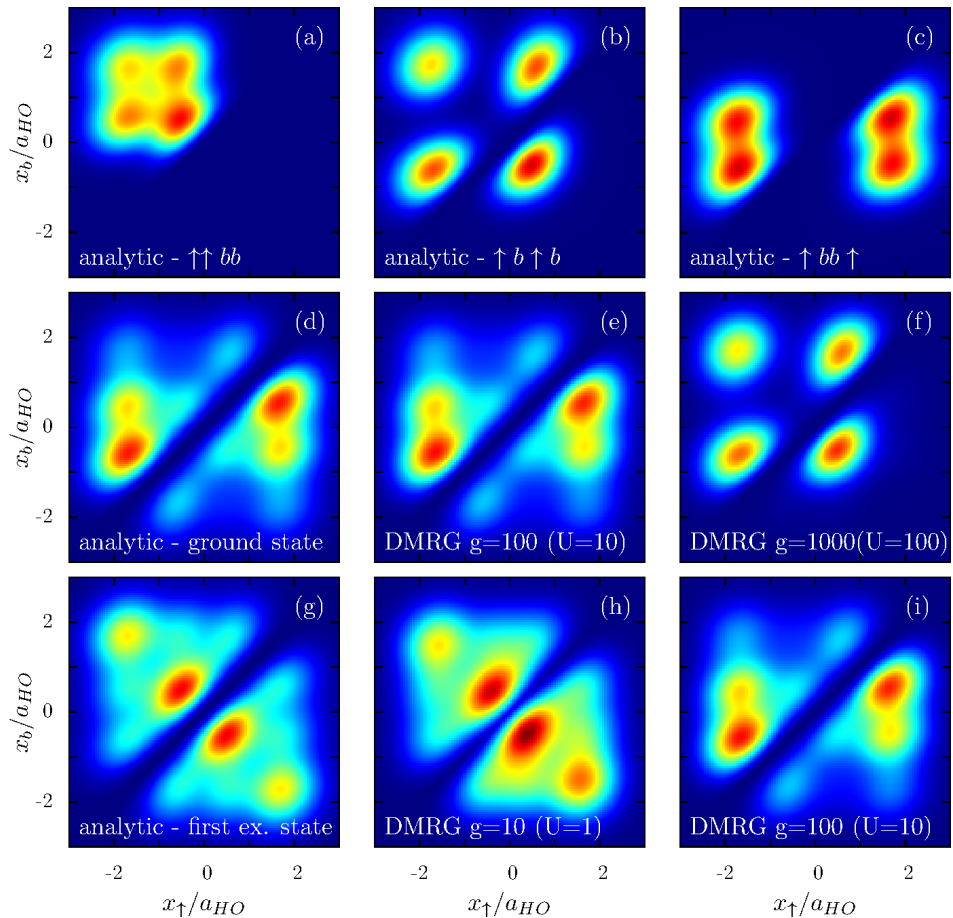}%
\caption{Pair correlation functions of a $N_\uparrow=N_b=2$ system. Panels (a) to (c) show the analytical results respectively for the spatial configurations $\uparrow \uparrow bb$, $\uparrow b \uparrow b$ and $\uparrow b b \uparrow$. Panels (d) and (g) give analytical results for ground and first excited states. DMRG results for the ground state are presented in panels (e) and (f) for the first excited state in panels (h) and (i). The parameters used for the DMRG calculations are as given in the
caption of figure~\ref{gs22}.}%
\label{pc22}%
\end{figure*}

To illustrate the spatial configurations it is advantageous to consider
the pair correlation function between the bosons and the fermions which
contains information on their relative positions in the harmonic trapping
potential. The pair correlation is given by
\begin{equation}
\textrm{P}\left(x,y\right) = 
\int dx_1...dx_N \delta(x-x_1)~\delta(y-x_N) \left| \psi(x_1,...,x_N) \right|^2.
\label{paircoreq}
\end{equation}
In the present case we have $N=4$ and since there are two different kinds of particles, 
the particles that correspond to coordinates $x_1$ and $x_4$ will be a boson and a fermion, 
respectively, as can be verified by considering the analytical form of the wave function 
in Eq.~\eqref{psiana}. We can therefore also refer to this pair correlation as the 
boson-fermion correlation function. 
Some examples of spatial configurations are shown in figure~\ref{pc22}(a), (b) and (c). 
The panels show spatial configurations $\uparrow \uparrow bb$  (associated to 
the amplitude $a_1$ in Eq.~\eqref{psiana}) in panel (a), $\uparrow b \uparrow b$  (coefficient $a_2$) in panel (b) 
and $\uparrow b b \uparrow$ (coefficient $a_3$) in panel (c). In panel 
(a) we see the pair correlation focused in the upper left-hand corner
where boson-boson and fermion-fermion distances are small, while the 
boson-fermion distance has two maximum that are rather close still, and 
similarly for panels (b) and (c). Notice that the configurations 
$b \uparrow \uparrow b$ (coefficient $a_4$), $b \uparrow b \uparrow $ (coefficient $a_5$) and
$b b \uparrow \uparrow $ ($a_6$) are not shown but may be very easily obtained 
by switching bosons and fermions which corresponds to a reflection in the $x_b=x_f$
plane in panel (a), (b) and (c) in figure~\ref{pc22}. 

Let us first consider the analytical wave functions in the strongly
interacting limit. The amplitudes of the different spatial 
configurations as given in Eq.~\eqref{psiana}
are 
\begin{align*}
&(a_1,a_2,a_3,a_4,a_5,a_6) = \\
&(−0.222, 0.448, −0.669, −0.226, 0.448, −0.222),
\end{align*}
for the ground state and 
\begin{align*}
(a_1,a_2,a_3,a_4,a_5,a_6)=(0.5, -0.5, 0.0, 0.0, 0.5, -0.5),
\end{align*}
for the first excited state.
The combination of the six spatial configurations properly weighted by their coefficients $a_k$ gives the pair correlation function shown in Figure~\ref{pc22}(d) for the ground state and in Figure~\ref{pc22}(g) for the first excited state (the respective densities can be seen in Figure~\ref{d22}). For the ground state, we see
that the dominant configuration corresponds to amplitude $a_3$, i.e to 
the two bosons sitting in the center and the fermions on the edge of the
trap as seen in the densities of figure~\ref{d22}. However, there are 
considerable contributions from all the other configurations as well. 
In the pair correlation function this is reflected since panel (d) for 
the ground state has its dominant non-zero contributions in the same
places as that seen for the configuration corresponding to coefficient
$a_3$ as seen in panel (c). The difference comes from a large admixture
of the configuration in panel (b) and a smaller contribution from that in 
panel (a). The first excited state is much more 'clean' in the sense 
that it only has configurations corresponding to coefficients $a_1$ and 
$a_2$, and with opposite signs. Panel (g) in figure~\ref{pc22} therefore
corresponds to a superposition of the results shown in panel (a) and (b). 

We note that these results also show that there is no simple relation among
the coefficients of the configurations in general. For instance, if we consider
a system of three equal mass particles where two are identical fermions, an 
allowed state for $1/g=0$ is 
\begin{align*}
&\psi_{(2+1)}(x_1, x_2, x_3) = \\
&(x_1-x_2) \cdot | x_2-x_3| \cdot | x_1-x_3 | \cdot \exp(-x_1^2-x_2^2-x_3^2),
\end{align*}
where $x_1$ and $x_2$ are the coordinates of $N_{\uparrow}$ and $x_3$ the coordinate of $N_{\downarrow}$. 
This wave function is inspired by a Tonks-Girardeau state which can be written as a 
Vandermonde determinant \cite{GirardeauPRA2004}.
But this is not a ground state and does not yield a state that is adiabatically connected
to other states in the spectrum for large but finite interaction strength \cite{VolosnievNC2014}. The 
relation of the exact solutions and the wave functions obtained from these Vandermonde inspired 
wave functions has been discussed in detail in Ref.~\cite{LevinsenSA2015}. Note that 
if we are discussing two identical bosons and a third particle, then making the 
substitution $(x_1-x_2)\to|x_1-x_2|$ in the wave function above will in fact give us 
a valid ground state as long as all two-body interactions are equal and large as it 
becomes the Girardeau wave function \cite{ GirardeauJoMP1960} which is the ground 
state also for two-component bosons with equal interactions between all pairs \cite{VolosnievPRA2015}.

\begin{figure*}[!htb]%
\includegraphics[width=\linewidth]{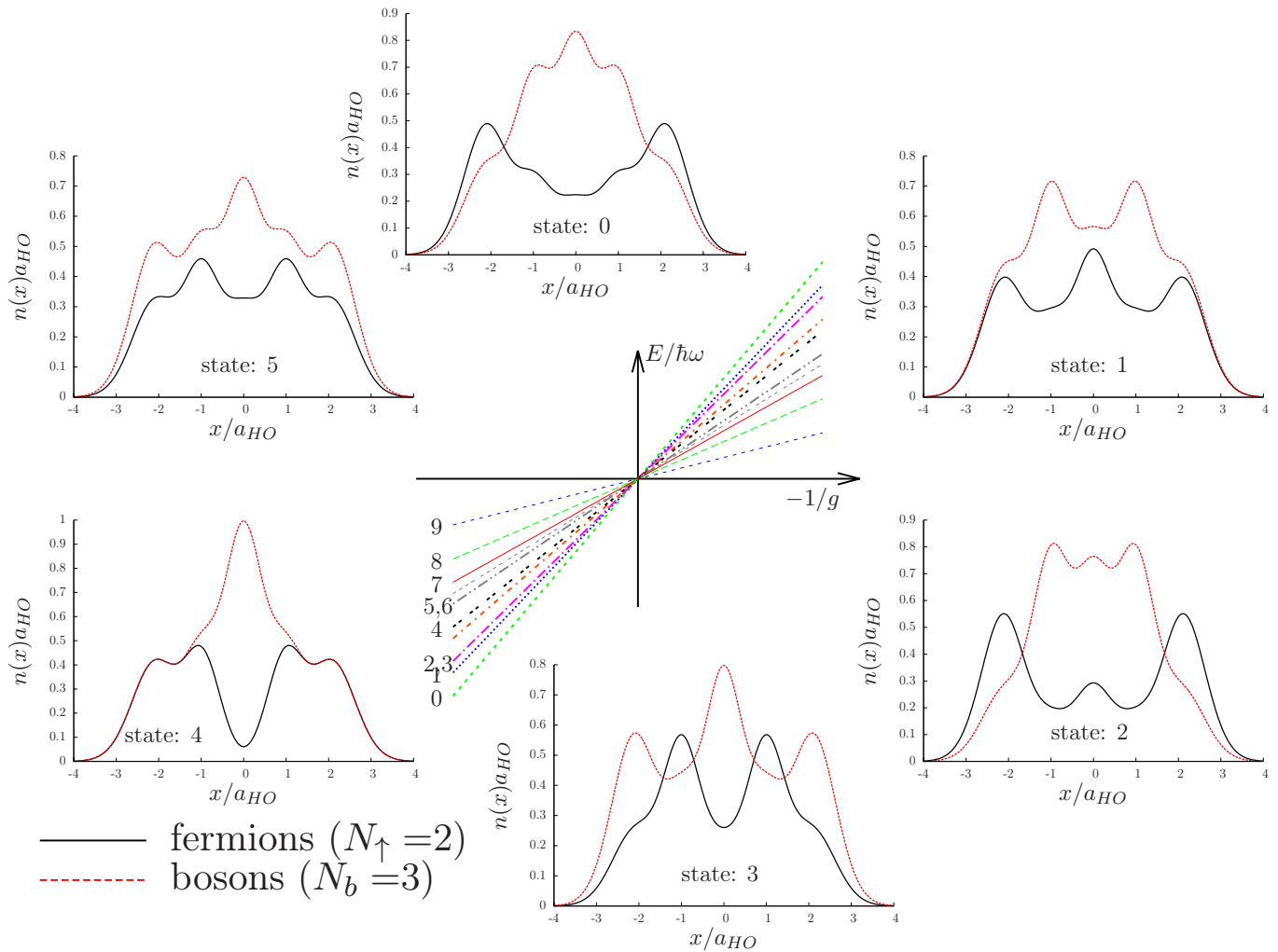}%
\caption{Energy slopes for the 10 degenerated states at $1/g=0$ and density profiles for the six lowest lying states for a mixture of 2 fermions and 3 bosons. Results obtained with the wave function Eq.~\ref{psiana} following the procedure from \cite{VolosnievNC2014}.}%
\label{d23}%
\end{figure*} 

In closing this section, we compare the pair correlation functions obtained from 
the analytical methods to the DMRG results. For the ground state, 
these are shown in 
panel (e) and (f) in figure~\ref{pc22}. Here we see that there is quite a good
resemblance of the DMRG pair correlation results for $U=10$ and that for the 
analytical results in panel (d). This can be immediately contrasted with 
the results from DMRG for $U=100$ as shown in panel (f) which clearly looks
nothing like the ground state. Curiously, it looks instead more or less 
perfectly like the results shown in panel (b) in figure~\ref{pc22}. This 
would lead us to conclude that the DMRG predicts that the wave function 
is that of the single spatial configuration with structure $\uparrow b \uparrow b$
which is of course not the case. Changing slightly the parameters and 
initialization of the DMRG routine may prompts the system to get stuck in 
some other spatial configuration as all of these become degenerate for 
extremely large interactions such as $U=100$ corresponding to $g\sim 1000$. 
For the first excited state the DMRG results are shown in panel (h) of 
figure~\ref{pc22} and this looks very similar to the analytical results 
in panel (g). However, notice here that we have a quite small $U=1$ and 
thus $g\sim 10$. For larger $U$ we again get stuck in the wrong kind 
of state as shown in panel (i), this time in something that looks very similar 
to the ground state wave function as given in panel (d). We find that for 
higher excited states one should generally expect that the strongly 
interacting limit sets in at smaller values of $U$ as compared to the 
ground state, see also \cite{BellottiPJD2017}.

\subsection{Larger systems}
Next, we analyze mixtures with a higher number of particles in 
order to look for some trends in larger systems.
We first analyze mixtures with $N_\uparrow=2$ and $N_b=3$, which have $M=5!/(2!3!)=10$ degenerated states at $1/g=0$. 
This is an odd-even mixture similar to the 
case highlighted in Ref.~\cite{deuretzbacher2016spin}, although here
for a smaller total number of particles. The 
results for the energies to linear order in $1/g$ near $1/g=0$ 
are shown in figure~\ref{d23} as obtained from Eq.~\eqref{psiana}. 
Along the edge of the figure we show the density profiles of a 
subset of the states. 

\begin{figure*}[!htb]%
\includegraphics[width=\linewidth]{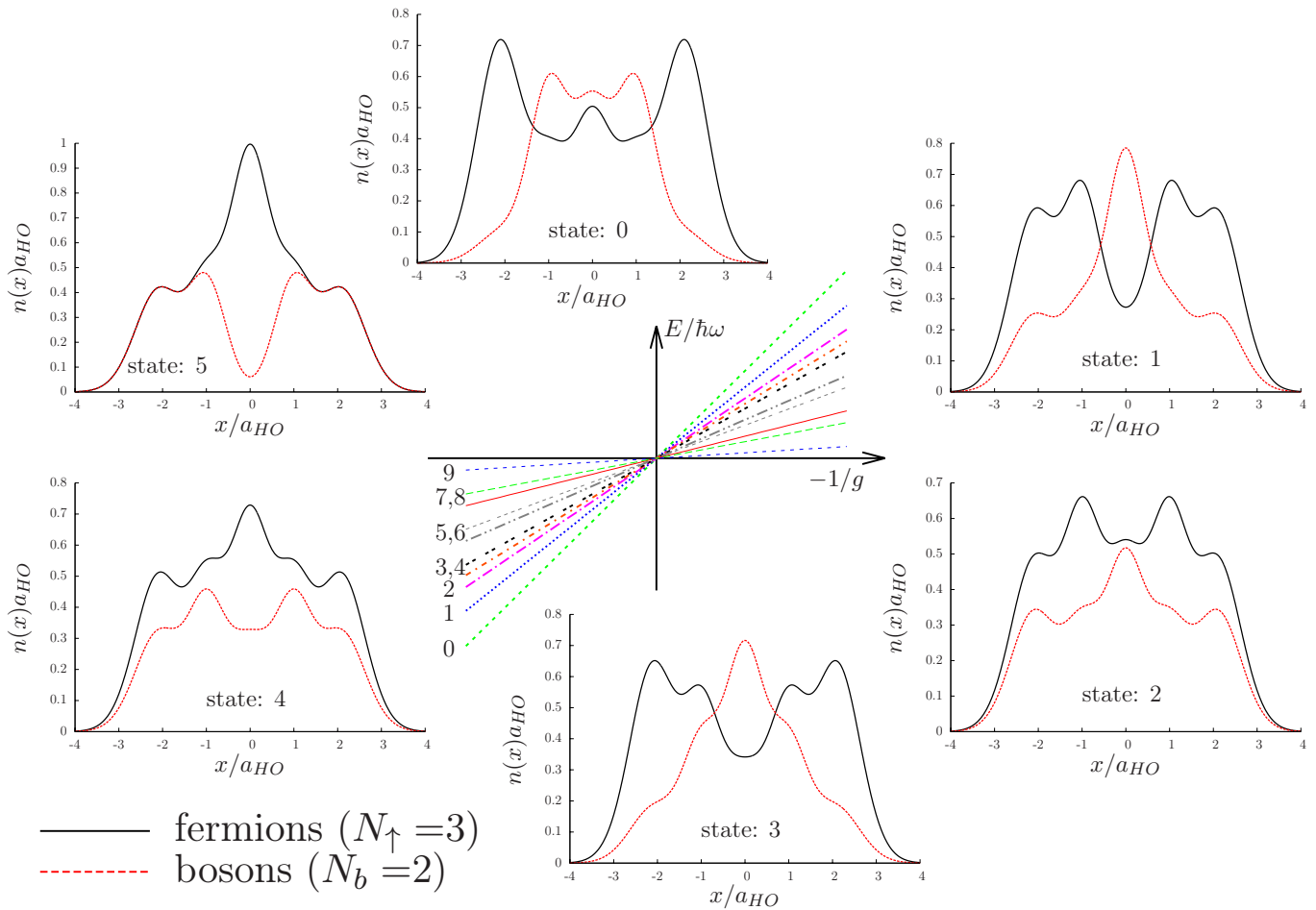}%
\caption{Energy slopes for the 10 degenerated states at $1/g=0$ and density profiles for the six lowest lying states for a mixture of 3 fermions and 2 bosons. Results obtained with the wave function Eq.~\ref{psiana} following the procedure from \cite{VolosnievNC2014}.}%
\label{d32}%
\end{figure*} 

The density profiles for the first six states in figure~\ref{d23} 
now become more elaborated and some states have the bosons and 
fermions tending to intercalate (sit in between each other). 
This is very different from the case with $N_\uparrow=N_b=2$
above, where there are two kinds of configurations, either 
species in the center and the other on the side, or the 
'democratic' fermionization corresponding to a density akin 
to four identical fermions. This latter kind of order is not 
seen in any of the ten states for the five-body case. 
A state that is somewhat like the democratic choice is that
denoted state 4 in figure~\ref{d23} which corresponds roughly 
to a boson occupying the center alone, and then on the wings
we have equal amounts of bosons and fermions. Note also that 
we do not find state where the fermions dominate the occupancy
in the center region. The ground state and state 2 are seen 
to have separated densities (akin to ferromagnetism), while 
states 1 and 3 have alternating boson-fermion structure. 
In particular, state 3 seems to carry a strongly antiferromagnetic 
tendency. The ground state profile seen here is consistent with that 
found in the odd-even $N=17$ case studied in Ref.~\cite{deuretzbacher2016spin}.

The next case we consider is one in which we still have five particles
but this time with $N_\uparrow=3$ and $N_b=2$. Again, there are 
$M=10$ states that become degenerate in energy at $1/g=0$. The 
results for this system are shown in figure~\ref{d32}. 
The linear slopes of the energy around $1/g=0$ are only 
slightly different from the other five-body case above.
However, the ground state profile is now very different 
as it shows clear intercalation between the bosons 
and fermions, while with a majority of bosons above 
we saw the bosons push the fermions out of the center of 
the harmonic trap.
Interestingly, we see that the states denoted 2 and 4 in figure~\ref{d32} 
are very similar in their structure if one interchanges the bosons with fermions. However, the other profiles are completely different. As before, systems composed of $N_\uparrow=3$ and $N_b=2$ do not possess any democratically fermionized state resembling that
of five identical fermions in their density profiles.

The discussion of the balanced four-body system in the previous 
section and the two ways to imbalance the system in the five-body 
case lead is to conclude that even in the case of these relatively modest 
size systems one may have very pronounced odd-even effects in 
Bose-Fermi mixtures.

\begin{figure*}[!htb]%
\includegraphics[width=\linewidth]{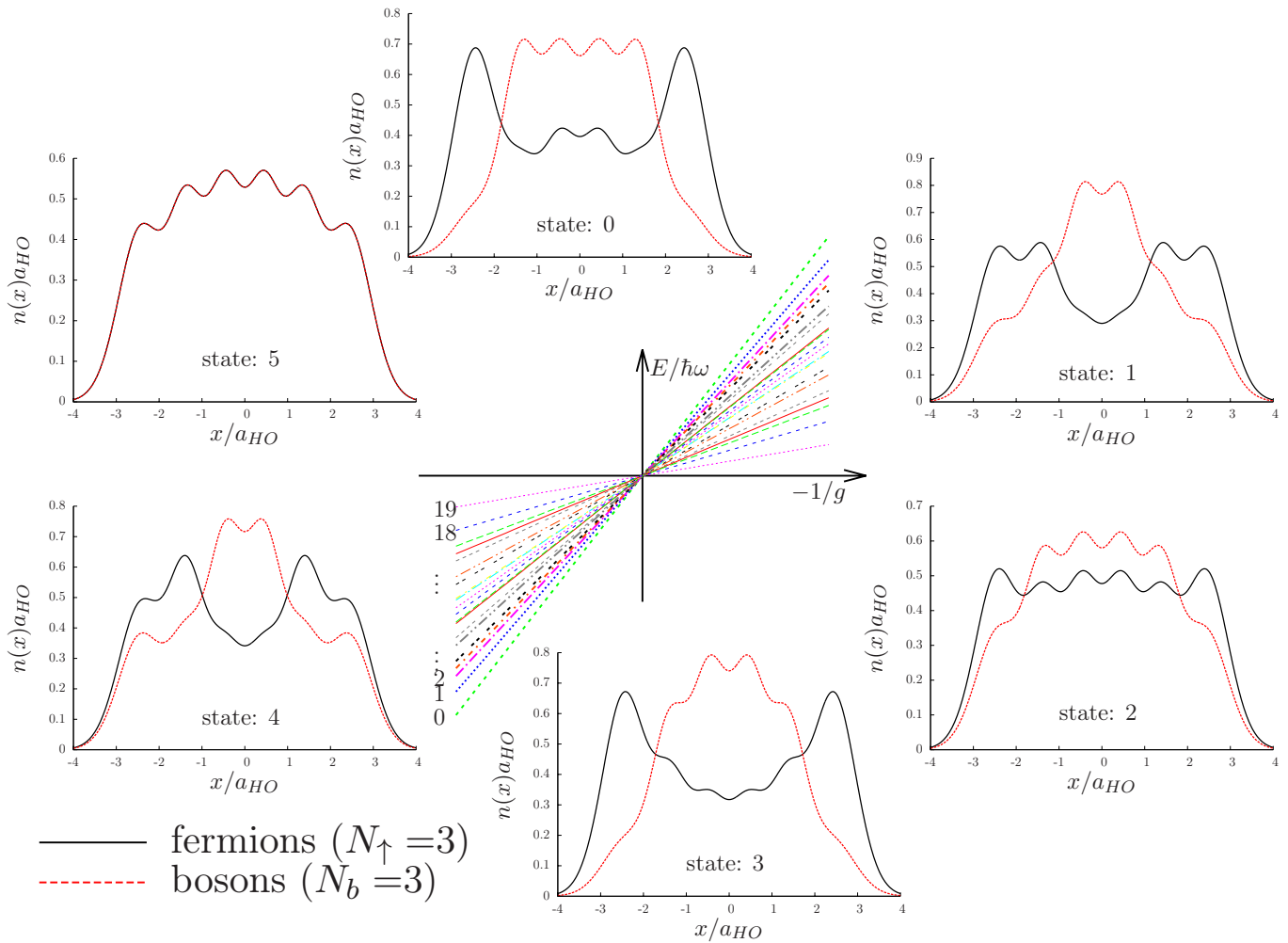}%
\caption{Energy slopes for the 20 degenerated states at $1/g=0$ and density profiles for the six lowest lying states for a mixture of 3 fermions and 3 bosons. Results obtained with the wave function Eq.~\ref{psiana} following the procedure from \cite{VolosnievNC2014}. }%
\label{d33}%
\end{figure*}

The last system we consider is a balanced setup of fermions and bosons 
with $N_\uparrow=N_b=3$. In this case we have $M=20$ degenerate states at $1/g=0$.
The slope of the energy around $1/g=0$ for all these states are shown in figure~\ref{d33}
along with the density profiles for the six lowest lying states. 
In the ground state, bosons are again pushed to the center of the trap 
and two fermions are clearly sitting at the edge of the trap. 
However, a third fermion seems 
to be sticking around between the bosons close to the origin. 
This time we do see the presence of a democratically fermionized
state which is the sixth excited state for which the density 
is also plotted in figure~\ref{d33}. We also see some rather peculiar
mixed density profiles in the manifold of excited states. Consider
for instance the state denoted by 2 in figure~\ref{d33} which seems
to have all the bosons centered around the middle, but at the 
same time similar peaks in the fermionic density. This could indicate
that a single fermion is sitting at the edge while the other 
two fermions are 'democratically' mixing with the bosons in 
the center. In comparison to the four- and five-body cases
we do see some similar trends, and again we see strong 
odd-even effects when going from five to six particles.

\section{Conclusion}\label{sec.conc}

We studied Bose-Fermi mixtures of four, five and six particles in 
both balanced and imbalanced systems. The ground state of four and 
six particles systems with equal number of bosons and fermions turns out
to try to separate the two kinds, bosons in the middle of the trap and 
fermions out on the wings. For imbalanced systems with five particles, we 
find that the particles would rather like to blend together, i.e. they
seem to be miscible in imbalanced systems. 
This suggests that phase 
separation of Bose-Fermi mixtures is an extremely delicate question 
in the few-body case \cite{ZinnerPRA2015}. In fact, even the density profile of the ground state
depends sensitively on whether it is the bosons or fermions that are 
present in an odd number. Previous studies in large homogenous systems 
\cite{DasPRL2003,CazalillaPRL2003,ImambekovPRA2006} also suggest that the question of 
whether the bosons and fermions mix or not is very delicate and 
depends on the theoretical approach. The study in Ref.~\cite{ImambekovPRA2006}
does find good agreement with our results in the few-body limit by using a 
local density approximation for the harmonic trap. 
A study using density-functional 
theory in a harmonic trap \cite{WangPRA2012} seem to agree with our few-body
finding of separation for balanced systems, but also finds evidence for
separation (fermions pushed to the side of the trap) in imbalanced systems
where we find a larger tendency for mixing. In particular, for the case
with more fermions than bosons, we find that a fermion can penetrate 
the bosonic density, which seems not to be the case in Ref.~\cite{WangPRA2012}. 
This may become less pronounced for larger particle numbers and the
situation would be very interesting to study in the future to see
where the few-body character spreads out as function of system size.
Another very noteworthy feature of our study is that 
among the excited states for balanced systems with 
equal numbers of bosons and fermions 
'democratically' fermionized states appear. These states have densities 
identical to a system of identical fermions with the same total 
number of particles, but this is the case for both the fermionic and 
the bosonic density profile. This is, however, not the 
case for the five particle imbalanced systems that we have studied, 
and again indicates a pronounced odd-even effect in these systems.

Our focus here has been on repulsive interactions and we have not 
touched on the case with attractive interactions where one has additional 
(deeply) bound states in the spectrum. For the DMRG method this may be 
rather difficult to describe but should in principle be possible. 
For the analytical results at very large interaction strength there is 
an interesting mapping from the repulsive to the attractive side 
as recently pointed out by \cite{deuretzbacher2016spin} in the context
of effective spin chain models for the strongly interacting regime. 
The statement is that upon changing the sign of the interaction while 
simultaneously changing bosons for fermions will only change the spin 
chain Hamiltonian by an overall sign. This implies that the corresponding 
spectrum is inverted. The presence of this mapping in our results can 
be seen in several places. In the balanced 2+2 system in figure~\ref{d22}
this is seen by comparing states 0 and 5, 1 and 4, as well as 2 and 3, all 
of which have identical densities if bosons and fermions are interchanged. The 
same is true for the 3+3 case although we have not displayed the full spectrum 
here. For the imbalanced cases with five bodies, one should compare states
4 and 5 in figures~\ref{d23} and \ref{d32} which are seen to be the same
under exchange of fermions and bosons. In this way, we can already infer
information about the strongly attractive regime from our data, although 
not for the deeper bound states that also arise with strong attraction. 

\acknowledgments
This work was supported by the Danish Council for Independent Research DFF
Natural Sciences, the DFF Sapere Aude program, and the Villum Kann Rasmussen
foundation. The authors thank M.~E.~S. Andersen, N.~J.~S. Loft, A.~S. Jensen, 
D.~V. Fedorov, M. Valiente and U. Schollw{\"o}ck for discussions, and in 
particular we thank F. Deuretzbacher for extended discussions and for 
pointing out a missing factor 
of half in our bosonic interaction terms. Finally, we thank D. Pecak for discussion 
of results and comparison of our calculations to results obtained using 
exact diagonalization.

\end{document}